\documentclass[12pt]{article}
\usepackage{times}  
\usepackage{helvet} 
\usepackage{courier}  
\usepackage[hyphens]{url}  
\usepackage{graphicx} 
\usepackage{graphicx}  
\usepackage{amsmath}
\usepackage{booktabs}
\usepackage{algorithm}
\usepackage{amssymb}
\usepackage{subfig}
\usepackage{array}
\usepackage{multirow}
\usepackage{placeins}
\usepackage{natbib}
\usepackage{hyperref}
\renewcommand{\cite}{\citep}
\usepackage{bm}
\usepackage{enumitem}
\usepackage{extarrows}
\usepackage[american]{babel}
\usepackage{pifont} 

\usepackage[ruled,noresetcount,algo2e]{algorithm2e}
\newenvironment{sciabstract}{%
\begin{quote} \baselineskip14pt\small\hfil {\bf Abstract} \hfil\\[3pt]}
{\end{quote}\vspace{6pt}}

\usepackage{algorithm}
\usepackage{algorithmic}
\usepackage{amsfonts} 
\usepackage{amsmath} 

\DeclareMathOperator*{\Adam}{Adam}

\newtheorem{assumption}{Assumption}

\newcounter{lastnote}

\title{Symbolic Cognitive Diagnosis via Hybrid Optimization for Intelligent Education Systems}

\author{
    Junhao Shen,
    Hong Qian$^*$,
    Wei Zhang,
    Aimin Zhou\\
\normalsize{
 Shanghai Institute of AI for Education and School of Computer Science and Technology
}\\
\normalsize{
East China Normal University, Shanghai 200062, China
}\\
\normalsize{
shenjh@stu.ecnu.edu.cn, \{hqian, amzhou\}@cs.ecnu.edu.cn, zhangwei.thu2011@gmail.com
}\\
\normalsize{$^*$Corresponding Author}
}

\date{}

\begin{document}

\baselineskip16pt

\maketitle 

\begin{sciabstract}
Cognitive diagnosis assessment is a fundamental and crucial task for student learning. It models the student-exercise interaction, and discovers the students' proficiency levels on each knowledge attribute. In real-world intelligent education systems, generalization and interpretability of cognitive diagnosis methods are of equal importance. However, most existing methods can hardly make the best of both worlds due to the complicated student-exercise interaction. To this end, this paper proposes a symbolic cognitive diagnosis~(SCD) framework to simultaneously enhance generalization and interpretability. The SCD framework incorporates the symbolic tree to explicably represent the complicated student-exercise interaction function, and utilizes gradient-based optimization methods to effectively learn the student and exercise parameters. Meanwhile, the accompanying challenge is that we need to tunnel the discrete symbolic representation and continuous parameter optimization. To address this challenge, we propose to hybridly optimize the representation and parameters in an alternating manner. To fulfill SCD, it alternately learns the symbolic tree by derivative-free genetic programming and learns the student and exercise parameters via gradient-based Adam. The extensive experimental results on various real-world datasets show the superiority of SCD on both generalization and interpretability. The ablation study verifies the efficacy of each ingredient in SCD, and the case study explicitly showcases how the interpretable ability of SCD works.
\end{sciabstract}

\section{Introduction}
Cognitive diagnosis assessment (CDA)~\cite{liu2021toward,liu2023new} is a fundamental task in intelligence education systems~\cite{AndersonHKL14,burns2014intelligent} with plenty applications, such as educational recommendation systems~\cite{xu2020recommendation} and exercise design~\cite{jeckeln2021face}. An illustrative example of CDA is shown in Figure~\ref{fig::CDAExample} of Appendix. There are three main factors in CDA: students, exercises and knowledge attributes which are also referred to as skills (e.g., calculation). The purpose of CDA is to model the student-exercise interaction via an interaction function based on response logs, and diagnose the students' cognitive states, i.e., inferring the proficiency levels on knowledge attributes.

In the past decades, extensive efforts have been dedicated to developing CDA methods. To name a few, item response theory (IRT) \cite{IRT}, multidimensional IRT (MIRT) \cite{MIRT}, deterministic inputs, noisy and gate model (DINA)~\cite{DINA}, neural cognitive diagnosis model (NCDM)~\cite{NCDM}, knowledge-association neural cognitive diagnosis model (KaNCD)~\cite{NeuralCD}, and Q-augmented causal cognitive diagnosis model (QCCDM)~\cite{liu2023qccdm}.
Although these cognitive diagnosis models (CDMs) have achieved remarkable progress, applying them to educational scenarios still suffers from the dilemma of generalization and interpretability. In real-world education scenarios, generalization (e.g., prediction of students' responses) and interpretability (e.g., interaction function and proficiency levels) are of equal importance for evaluating a cognitive diagnosis method~\cite{hassan2022explainable}. The procedure and outcome of a CDM should be comprehensible and trustworthy for users such as students, teachers and parents. However, due to the complicated student-exercise interaction~\cite{dibello200631a}, most existing methods can hardly make the best of both worlds.

Specifically, on the basis of the $\rm{sigmoid}$ interaction function, models like IRT and MIRT exhibit the interpretability of interaction functions and relatively good generalization, but the diagnostic outcomes (e.g., proficiency levels and exercise difficulty parameters) lack interpretability due to the latent vectors vaguely corresponding to each knowledge attribute~\cite{embretson2013item}. Under the conjunctive assumption, models like DINA possess the interpretability of both interaction function and outcomes, but they may underperform in terms of generalization due to their simple forms of interaction function~\cite{de2011generalized}. Through neural networks, models like NCDM show the strong generalization and interpretability of outcomes, but they could lack interpretability of the interaction function due to the black-box nature of neural networks~\cite{murdoch2019definitions,du2020techniques}.
Most existing methods struggle to well balance the generalization and interpretability mainly because of the dilemma of accurately modelling the complex interaction function in a non-linear way and its intelligibility. 
To alleviate this dilemma, a highly non-linear and explicable representation of interaction function, symbolic regression~(SR), could provide a good recipe to make the best of both worlds. SR excels in finding a complicated non-linear function with high interpretability thanks to the tree-structured expression.

Unfortunately, directly adapting SR to CDA is infeasible due to the following issues arising from education. Firstly, although SR excellently performs in regression tasks~\cite{billard2002symbolic}, CDA is not only a regression problem but also a complex task that requires simultaneously obtaining the interaction function and diagnostic outcomes. Secondly, the interaction function achieved via SR could not satisfy the monotonicity assumption~\cite{reckase2009} which is vital common sense in education.

To this end, this paper proposes a symbolic cognitive diagnosis (SCD) framework to simultaneously boost interpretability of outcomes and interaction functions, while maintaining competitive generalization performance. The SCD framework employs the symbolic tree to explicably represent complicated student-exercise interaction functions, and utilizes gradient-based optimization techniques for effective learning of student and exercise parameters. At the same time, the accompanying challenge arises in reconciling discrete symbolic representation learning with continuous parameter optimization to enable model training. To address this challenge, we propose to hybridly optimize the representation and parameters in an alternating manner. To fulfill SCD, it alternately learns the symbolic tree by derivative-free genetic programming~\cite{o2009riccardo} and learns the student and exercise parameters via gradient-based Adam~\cite{kingma2017adam}, resulting in the SCD model~(SCDM). 
Specifically, preliminary student and exercise parameters are obtained via optimizing the manually designed initial interaction function. Then, these diagnostic outcomes are fixed to optimize the interaction function via symbolic regression. Afterwards, this complex interaction function is fixed for optimizing the parameters to obtain new diagnostic outcomes, and this process is repeated alternately. To satisfy the monotonicity assumption in education, the function set of symbolic regression only includes monotonic operators and the forms of trees are also subject to the constraints. 
The extensive experiment results on various real-world datasets show the excellence of SCD on both generalization and interpretability. The ablation study verifies the efficacy of each components of SCD. The case study explicitly showcases how the interpretable ability of SCD works in education scenarios, especially the interpretable form of the learned interaction function.

In the subsequent sections, we respectively recap the related work, introduce the preliminaries, present the proposed SCD, show the experimental results and analysis and finally conclude the paper.

\section{Related Work}
\textbf{Cognitive Diagnosis Assessment.} CDA is one of the most important research areas in educational psychology, with many representative CDMs. IRT~\cite{IRT} models the student' response to a exercise as an outcome of the interaction between the latent traits of student and the traits of exercise. Specifically, the $\rm{sigmoid}$ function containing the traits gives the probability of student correctly answering the exercise. In the following decades, these artificially designed interaction functions, such as MIRT~\cite{MIRT,reckase2009} and DINA~\cite{DINA}, have demonstrated increasing performance in CDA through parameter expansions~\cite{fischer1995derivations} and increased dimensions~\cite{MIRT,chalmers2012mirt}. However, the weakness also becomes obvious: the manually designed interaction functions hardly fully capture the complex student-exercise relationship~\cite{reckase2009}.


Therefore, CDMs based on artificial neural networks have emerged. Given that neural networks have been theoretically demonstrated to possess the universal approximation property~\cite{hornik1989multilayer}, they can be employed to effectively capture the intricate student-exercise interactions in the CDA, replacing the traditional interaction function. Specifically, in NCDM~\cite{NCDM}, KaNCD~\cite{NeuralCD} and QCCDM~\cite{liu2023qccdm}, the neural networks are designed as the multiple full connection layers to capture the complex student-exercise interactions. 
Nevertheless, interpretability is also of equal importance in education scenarios~\cite{hassan2022explainable}, leading to the insufficiency of those black-box neural networks.

\textbf{Symbolic Regression.} Symbolic regression~(SR) is an effective approach to finding a suitable mathematical model to describe data~\cite{billard2002symbolic}. Compared with traditional regression techniques like linear regression, there are no a priori assumptions on the specific form of the function. In other words, SR explores through the mathematical expression spanned by candidate mathematical operators to discover the most suitable solution~\cite{zhang2022ps}. SR plays a significant role in data processing and analysis across various fields due to its interpretability, e.g., dynamical systems prediction~\cite{quade2016prediction}. There are also some educational applications like academic performance prediction~\cite{fan2023integration}. However, adapting SR to CDA remains challenging since CDA is not merely a prediction task. Currently, to the best of our knowledge, relevant work that adequately applies SR to CDA seldom exists. In this paper, we propose a SCD framework which balances the generalization and interpretability of both interaction function and diagnostic outcomes, especially the interpretable form of the learned interaction function.

\section{Preliminaries}
\textbf{Task Overview.} We at first introduce some notations and definitions in cognitive diagnosis. Let $S=\{s_1, \dots, s_N\}$, $E = \{e_1, \dots, e_M \}$ and $K = \{k_1, \dots, k_L\}$ respectively denote the sets of students, exercises and knowledge attributes, where $N=|S|$, $M=|E|$ and $L=|K|$ are the size of each set. Each student is required to finish some exercises for practice, and the corresponding response logs are denoted as a set of triplets $R=\{(s, e, r)|s \in S, e \in E, r \in \{0, 1\}\}$, where $r$ represents the score (i.e., $r=1$ means right while $r=0$ means wrong) obtained by student $s$ on exercise $e$. Besides, the Q-matrix (usually annotated by experts) is denoted as $\bm{Q} = \{Q_{i,j}\}_{M \times L}$, which captures the relationship between exercises and knowledge attributes. Herein, $Q_{i,j} = 1$ if exercise $e_{i}$ is associated with knowledge attribute $k_j$, and $Q_{i,j} = 0$ otherwise. Furthermore, the monotonicity assumption~\cite{reckase2009}, as Assumption~\ref{def::ma}, is introduced to enhance the interpretability of diagnostic outcomes.

\begin{assumption}[Monotonicity Assumption]
\label{def::ma}
The probability of correctly answering an exercise is monotonically increasing at any dimension of the student's proficiency level on relevant knowledge attributes.
\end{assumption}

\textbf{Problem Definition.} Given the observed triplet logs of students $R$ and the labelled Q-matrix $\bm{Q}$, the goal is to infer the students’ proficiency on knowledge attributes with the process of simultaneously estimating the interaction function and proficiency level, where the interaction function models the student-exercise interaction.

\textbf{Genetic Programming.} Genetic programming (GP)~\cite{koza1994genetic,yi2023expla} stands out as one of the most prevalent methods used in SR. It aims to evolve solutions to a given problem following Darwin's theory of evolution, seeking the fittest solution over several generations. Instead of using binary code to represent chromosomes as in genetic algorithms~\cite{forrest1993genetic}, solutions in GP are represented as tree-structured chromosomes containing nodes and terminals, forming a symbolic tree. The tree comprises interior nodes denoting mathematical operators and terminal nodes denoting variables. Employing the depth-first search, the mathematical expression for each individual solution can be obtained through traversing the tree. The GP procedure will be clarified in the SCD Model (SCDM) Implementation Section.

\section{Symbolic Cognitive Diagnosis (SCD)}
This section presents the key ingredients of SCD. The overall framework of SCD is shown in Figure~\ref{fig::overview}. Sequentially, we introduce the basic diagnostic factors of cognitive diagnosis in the SCD framework, elaborate the procedure of SCD framework, fulfill the SCD framework with GP and Adam optimizer to result in the proposed SCD model (SCDM), and finally discuss the flexibility of the SCD framework.

\subsection{Diagnostic Factors}
Generally, there are three factors in CDA need to be diagnosed and modeled, i.e., proficiency levels, exercise features and interaction function~\cite{dibello200631a}. Details are introduced as bellow.

\begin{figure*}[!t]
  \centering
  \includegraphics[width=\linewidth]{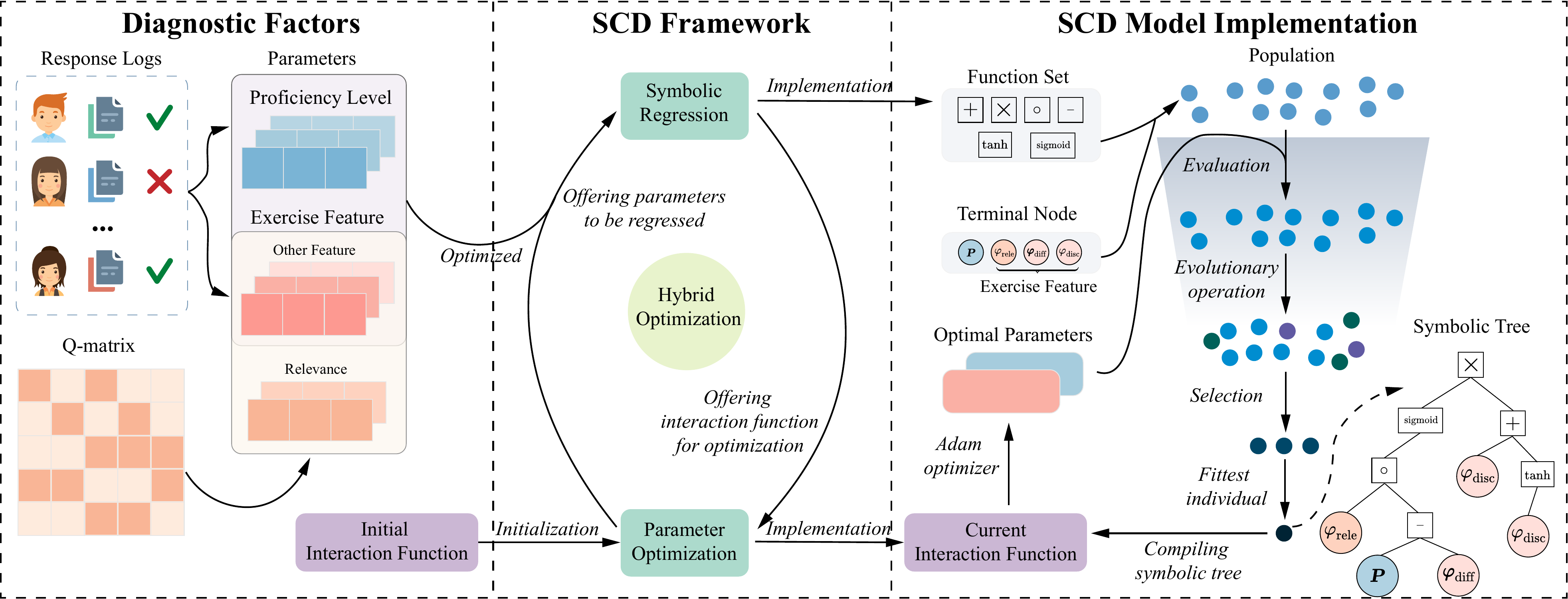}
  \caption{An overview of the proposed symbolic cognitive diagnosis (SCD)}
  \label{fig::overview}
\end{figure*}

\textbf{Proficiency Levels.} 
We aim to assess students' proficiency levels directly, whose each dimension corresponds to a specific knowledge attribute, without utilizing latent vectors like IRT or MIRT. Specifically, proficiency levels are denoted as $\bm{P} = \{P_{i,j}\} \in \mathbb{R}^{N \times L}$, and the $i$-th student's proficiency level is $\bm{P}_{i, \bullet}$, each entry of which is continuous among $[0, 1]$ and indicates the student's proficiency on a certain knowledge attribute. For instance, $\bm{P}_{i, \bullet}=[0.2, 0.5, 0.8]$ means the $i$-th student has a low mastery on the first knowledge attribute, middle on the second, and high on the last. Proficiency levels are deemed as the student parameter and learnt during the continuous parameter optimization module.

\textbf{Exercise Features.} Exercise features refer to the characteristics of each exercise. We divide exercise features into two categories. The first involves the relevance between the $j$-th exercise and knowledge attributes, and is denoted as $\bm{\varphi}_{\text{rele}_j} \in \{0, 1\}^{L}$, which is directly from the Q-matrix $\bm{Q}$ and not trainable during the parameter optimization. The second involves other optional trainable exercise parameters such as exercise difficulty and discrimination, and they can be included if necessary.

As shown in the left of Figure~\ref{fig::overview}, proficiency levels and other exercise features are deemed as parameters, which are learnt during the parameter optimization.

\textbf{Interaction Function.} The interaction function models the process of students completing exercises and getting the response results. In the SCD, we utilize SR to obtain the interaction function for several reasons. Firstly, SR effectively captures non-linear and complex student-exercise relationship since it does not rely on a predefined functional form. Secondly, the result of model is illustrated by the symbolic tree as Figure~\ref{fig::overview}, which exhibits high interpretability. Thirdly, the function set and terminal nodes can be expended to accommodate various real-world educational tasks. Formally, the output of SCDM can be formulated as
\begin{equation}
    \label{eq::if}
    y_f = \sigma(f(\bm{P}_{i, \bullet}, \bm{\varphi}_{\text{rele}_j}, \bm{\varphi}_{\text{others}_j}))\,,
\end{equation}
where $y_f$ is the probability of the $i$-th student correctly answering the $j$-th exercise given by interaction function $f$ compiled from symbolic tree, $\sigma$ denotes the activate function, and $\bm{\varphi}_{\text{others}}$ denotes other factors except $\bm{P}$ and $\bm{\varphi}_{\text{rele}}$.

However, the symbolic tree constructed by a unlimited function set in Eq.~\eqref{eq::if} hardly meets Assumption~\ref{def::ma}. Thus, we need to design tailored strategies to address this problem.

\subsection{Symbolic Cognitive Diagnosis Framework}
This section introduces the proposed SCD framework shown as the middle of Figure~\ref{fig::overview}. This framework can be divided into two modules: parameter optimization~(PO) and SR. Due to the inherent differences between continuous and discrete optimization processes, they usually cannot be synchronously combined. Thus, the hybrid optimization is incorporated to asynchronously unify the continuous and discrete optimization processes and begins with PO.

\textbf{Parameter Optimization.} The reason for choosing to start with PO is that the quality of initialization significantly influences the model's training performance, and its initialization is easier compared with SR. Specifically, although both the student-exercise interaction function and parameters of students and exercises are unknown, manually designed simple interaction functions like ${\rm sigmoid}$ have been shown to effectively approximate for an unobserved interaction function in the field of education~\cite{IRT,MIRT,reckase2009}. Conversely, the parameters lack suitable estimates, and random initialization is likely to lead symbolic regression astray. After initializing the interaction function in the PO module, the objective function is $\mathcal{L}_{r\in R}(y_f, r)$, where the $\mathcal{L}$ is loss function, $r$ is the label from $R$, and $y_f$ is defined in Eq.~\eqref{eq::if}. After parameter optimization, the optimal parameters are proficiency levels and trainable exercise features (e.g., difficulty) and sent to SR.

\textbf{Symbolic Regression.} After gaining the optimal parameters, SR can discover the potential interaction function between students and exercises. This module does not have access to the interaction function in PO module. Some may ask: if SR directly finds the interaction function identical to the PO module, how do we discover more complex interaction relationships? Note that there are complexity requirements of SR. Typically, we aim to identify a suitable function as the regression result from complex candidates, which often outperforms simple ones in generalization (validated in ablation study). After regression, the optimal function is sent to the PO module as the current interaction function.

\subsection{SCD Model (SCDM) Implementation}
This section introduces the implementation of the SCD framework as the right of Figure~\ref{fig::overview}. The PO is implemented by the Adam~\cite{kingma2017adam}, SR by the GP~\cite{billard2002symbolic}, and hybrid optimization by alternative optimization~(AO). The SCDM implementation can be divided into two phases: continuous optimization and discrete optimization. Algorithm~\ref{alg::scdm} shows the implementation of SCDM.


\begin{algorithm}[!t]
\caption{Symbolic Cognitive Diagnosis Model (SCDM)}
\label{alg::scdm}
\textbf{Input}: Response logs $R=\{s, e, r\}$, Q-matrix: $\bm{Q}$, and initial interaction function $f_{\text{init}}$.\\
\textbf{Parameter}: Maximum number of epochs $T$, generation $T_{\text{GP}}$, population size $V$, crossover rate $p_{\text{cr}}$, and mutation rate $p_{\text{mu}}$.\\
\textbf{Output}: Proficiency levels $\bm{P}$, exercise features $\bm{\varphi}_{\text{diff}}, \varphi_{\text{disc}}$, and fittest interaction function $\hat{f}$.
\begin{algorithmic}[1] 
\STATE Initialize the $\bm{P}, \bm{\varphi}_{\text{diff}}, \varphi_{\text{disc}}$; $\hat{f} \gets f_{\text{init}}, \bm{\varphi}_{\text{rele}_j} \gets \bm{Q}_{j, \bullet}$
\FOR {$t=1,2, \dots, T$}
\STATE 
$\bm{P}, \bm{\varphi}_{\text{diff}}, \varphi_{\text{disc}} \gets \Adam \mathcal{L}_{r\in R}(y_{\hat{f}}, r)$
\STATE Randomly initialize a population of interaction functions $F=\{f_1, f_2, \dots, f_{V}\}$
\FOR {$t_1 = 1, 2, \dots, T_{\text{GP}}$}
\FOR {$i = 1, 2, \dots, V$}
\IF{$i \% 2 = 1$ and $ {\rm{uniform}}(0, 1) < p_{\text{cr}}$}
\STATE $\{f_{i}, f_{i + 1}\} \gets \text{crossover}(f_{i}, f_{i + 1})$
\ENDIF
\IF {${\rm{uniform}}(0, 1) < p_{\text{mu}}$}
\STATE $f_{i} \gets \text{mutation}(f_{i})$
\ENDIF
\ENDFOR
\STATE Evaluate all individuals in $F$, $F \gets \text{selection}(F)$
\ENDFOR
\STATE $\hat{f} \gets f$, where $f$ is the fittest in $F$
\ENDFOR
\STATE \textbf{return} $\bm{P}, \bm{\varphi}_{\text{diff}}, \varphi_{\text{disc}}, \hat{f}$
\end{algorithmic}
\end{algorithm}

\textbf{Continuous Optimization.} The continuous optimization for learning the student and exercise parameters $\bm{P}$, $\bm{\varphi}_{\text{diff}}$ and $\varphi_{\text{disc}}$ is described in line 3 of Algorithm~\ref{alg::scdm}. The initial interaction function is inspired by MIRT and formulated as
\begin{equation}
    \label{eq::init}
    f_{\text{init}} = \bm{\varphi}_{\text{rele}_j} \circ (\bm{P}_{i, \bullet} - \bm{\varphi}_{\text{diff}_j}) \times \varphi_{\text{disc}_j}\,,
\end{equation}
where $\circ$ is inner product, $\bm{\varphi}_{\text{diff}_j} \in \mathbb{R}^{L}$ is the difficulty of the $j$-th exercise, and $\varphi_{\text{disc}_j} \in \mathbb{R}$ is the discrimination of the $j$-th exercise. Eq.~\eqref{eq::init} is simple and intuitive, but can correctly lead the optimization. The activate function $\sigma$ is $\rm{sigmoid}$ which is widely used in CDMs. To use the Adam optimizer, we formulate the loss function of SCDM as the cross entropy between the $i$-th output of interaction function $f$ denoted by $y_f^{(i)}$ and the corresponding label $r^{(i)}$, which is Eq.~\eqref{eq::loss} below.
\begin{equation}
\label{eq::loss}
\mathcal{L}_{r\in R}(y_f, r) = -\sum\limits_{i=1}^{|R|}(r^{(i)}\log y_f^{(i)} + (1-r^{(i)})\log (1-y_f^{(i)}))\,.
\end{equation}
After training Eq.~\eqref{eq::loss}, the parameters $\bm{P}$, $\bm{\varphi}_{\text{diff}}$ and $\varphi_{\text{disc}}$ are the current diagnostic outcomes, which are sent to the discrete optimization as the parameters of symbolic regression.

\textbf{Discrete Optimization.} The discrete optimization for learning the symbolic representation tree is shown in line 5 to 15 of Algorithm~\ref{alg::scdm}. The procedure begins with an initial population of randomly generated individuals which must adhere to specific criteria (e.g., ensuring the output to be scalar). Through random operators like crossover and mutation, the current population is evolved. Then, each individual's fitness is evaluated by metric like accuracy. Finally, individuals are selected as parents in a certain way, and their offspring become the next generation. This process continues until it reaches the maximum generation. The fittest individual is the new interaction function and sent to continuous optimization. More detailed implementation is depicted in Appendix~\ref{para::gpsetting}.

\textbf{Discussions.} We discuss some points regarding SCD and SCDM. (1) Flexibility. Proficiency levels $\bm{P}$ and Q-matrix $\bm{Q}$ are essential in the SCD framework. SCDM includes difficulty and discrimination, and other exercise features can also be included if necessary. The initial interaction function is restricted to including $\bm{P}_{i, \bullet} \circ \bm{\varphi}_{\text{rele}_j}$, ensuring each dimension of $\bm{P}_{i, \bullet}$ corresponding to a specific knowledge attribute. (2) Interpretability. SCD involves interpretability in two aspects: diagnostic outcomes via Assumption~\ref{def::ma} and interaction functions guaranteed by GP. GP provides explicit operations and explainable object functions, and SR trees optimized by GP are transparent. (3) Implementation. We choose Adam~\cite{kingma2017adam} for optimization due to its wide use in deep learning, fast convergence, and to ensure fair comparison with existing methods that use it. The SCD framework is generic and can be implemented differently. For example, SR can be realized by other algorithms like transformer~\cite{kamienny2022end}, and PO can be realized by evolutionary strategies~\cite{beyer2002evolution}.


\section{Experiments}
This section conducts extensive experiments on real-world datasets to answer the following crucial questions. The source code of SCDM is available at GitHub\footnote{\url{https://github.com/shinkungoo/SymbolicCDM}}.

\begin{itemize}
    \item \textbf{Q1:} How does SCDM perform when compared with existing CDMs in terms of generalization?
    \item \textbf{Q2:} How does SCDM perform when compared with existing CDMs in terms of interpretability?
    \item \textbf{Q3:} How do GP and gradient optimization contribute to the performance of SCDM respectively?
    \item \textbf{Q4:} How do hyperparameters influence SCDM?
    \item \textbf{Q5:} How does the interpretable ability of SCDM work in real-world educational scenarios?
\end{itemize}

\subsection{Experimental Setup}
\textbf{Dataset Description.} The experiments are conducted on four real-world datasets, i.e., Math1, Math2~\cite{liu2018fuzzy}, FracSub~\cite{wu2015cognitive} and NeurIPS2020~\cite{wang2020diagnostic}. Math1 and Math2 consist of response logs from high school students taking the final exams of their first and second senior years respectively. FracSub comprises of scores of middle school students on fraction subtraction objective problems. NeurIPS2020 containing two school years of students’ answers to mathematics questions from Eedi. Details of these datasets are shown in Table~\ref{tab::datasets}.



\begin{table}[!h]
  \centering
  \caption{Statistics of real-world datasets for experiments.}
  \resizebox{0.7\linewidth}{!}{
    \begin{tabular}{l|cccc}
    \toprule
    Datasets & Math1 & Math2 & FracSub & NeurIPS2020 \\
    \midrule
    \#Students              & 4209      & 3911      & 536       & 4129 \\
    \#Exercises             & 15        & 16        & 20        & 44 \\
    \#Knowledge Attributes  & 11        & 16        & 8         & 30 \\
    \#Response Logs         & 63135     & 62576     & 10720     & 66638 \\
    Average Correct Rate    & 0.5515    & 0.4880    & 0.5339    & 0.5450\\
    \bottomrule
    \end{tabular} }
  \label{tab::datasets}
\end{table}

\textbf{Baselines and State-of-the-Art Methods.} Over the past few decades, CDMs have been developing, and here we select some representative approaches for comparison.
\begin{itemize}
    \item IRT~\cite{IRT} is a classical CDM that employs a logistic-like interaction function.
    \item MIRT~\cite{MIRT} is an extended model of IRT, which uses multidimensional $\bm{\theta}$ and $\bm{b}$ to model the latent traits of students and exercises.
    \item DINA~\cite{DINA} is a CDM based on the conjunctive assumption, where proficiency levels are represented by two discrete values, i.e., $0$ and $1$.
    \item NCDM~\cite{NCDM} is a CDM that replaces the traditional interactive function with neural networks, which outperforms the traditional CDA methods on most datasets with interpretability of the diagnostic outcomes.
    \item KaNCD~\cite{NeuralCD} is a CDM based on the improvements of NCDM, which considers the implicit association between knowledge attributes, and reaches the state-of-the-art on most datasets.
\end{itemize}

\textbf{Generalization Metrics.} Assessing the performance of CDM can be challenging due to the difficulty in obtaining accurate proficiency levels of students. To overcome this, a widely accepted approach to evaluating them is through the prediction of students' test scores. Accordingly, similar to previous CDMs~\cite{NCDM}, we assess how close the model’s prediction (whether a student solves a question or not) is to the ground truth in the test set with classification metrics, i.e., accuracy, area under curve (AUC) and F1-score (F1). 
%
Besides, in the field of SR, the $R^2$ score is usually employed for evaluating symbolic trees~\cite{zhang2022evolutionary}. As aforementioned, however, the true student-exercise interaction function is unknown, rendering this metric unavailable.

\textbf{Interpretability Metric.} Generalization metrics are only one aspect of evaluating the performance of CDMs. On the other hand, interpretability metrics hold equal significance in the education scenario. The interpretability of interaction functions is often assessed by the choice of function set and tree depth. But there are no other CDMs that utilize SR, we solely quantitatively assess the interpretability of the diagnostic outcomes via the degree of agreement (DOA)~\cite{NCDM,NeuralCD}. Intuitively, if student $s_a$ has a greater accuracy in answering exercises related to $k_i$ than student $s_b$, the proficiency level of $s_a$ should be higher than that of student $s_b$ on the knowledge attribute $k_i$, i.e., $\bm{P}_{a,i} > \bm{P}_{b,i}$. The DOA of $k_i$ is defined as Eq.~\eqref{eq::doa}.
\begin{equation}
\label{eq::doa}
DOA(i) = \frac{1}{Z}\sum\limits_{a=1}^{N}\sum\limits_{b=1}^{N}\delta(\bm{P}_{a,i}, \bm{P}_{b,i})\sum\limits_{j=1}^{M}I_{j,i}\cdot\frac{J(j,a,b) \land \delta(r_{aj}, r_{bj})}{J(j,a,b)}\,,
\end{equation}
where $Z=\sum_{a=1}^{N}\sum_{b=1}^{N}\delta(\bm{P}_{a,i}, \bm{P}_{b,i})$, and $\bm{P}_{a,i}$ is the proficiency level of student $s_a$ on knowledge attributes $k_i$. $\delta(x, y) = 1$ if $x > y$ and otherwise $\delta(x, y) = 0$. $I_{ji} = 1$ if exercise $e_j$ contains knowledge attribute $k_i$ and otherwise $I_{ji} = 0$. $J(j, a, b) = 1$ if both student $s_a$ and $s_b$ completed exercise $e_j$ and otherwise $J(j,a,b)=0$. To evaluate the interpretability of diagnostic outcomes, we average the $DOA(i)$ on all knowledge attributes.

\textbf{Detailed Settings.} In the GP module implemented by DEAP~\cite{deap}, the population size $V$ is $200$, the number of generation $T_{GP}$ is $10$, the crossover and mutation rates are $0.5$ and $0.1$ respectively, the initial tree depth is $5$, and the selection method is tournament selection. To meet the Assumption~\ref{def::ma}, the function set is $\{+, \times, \circ, -, \tanh, \rm{sigmoid}\}$. Most of the setting is referred to~\cite{zhang2022evolutionary}, which shows to be effective in SR. 
In the continuous optimization module implemented by Pytorch~\cite{pytorch}, we set the learning rate of Adam~\cite{kingma2017adam} to be $0.002$, and initialize the feature parameters in the interaction function with Xavier normal initialization~\cite{xavier}. 
To evaluate performance, the size of test dataset is $0.2$, and all experiments are repeated independently with $10$ seeds.

\subsection{Experimental Results}
We conduct comprehensive experiments to answer the aforementioned questions. More details are in Appendix~\ref{para::exp}.

\begin{figure*}[!tbp]
  \centering
  \includegraphics[width=\linewidth]{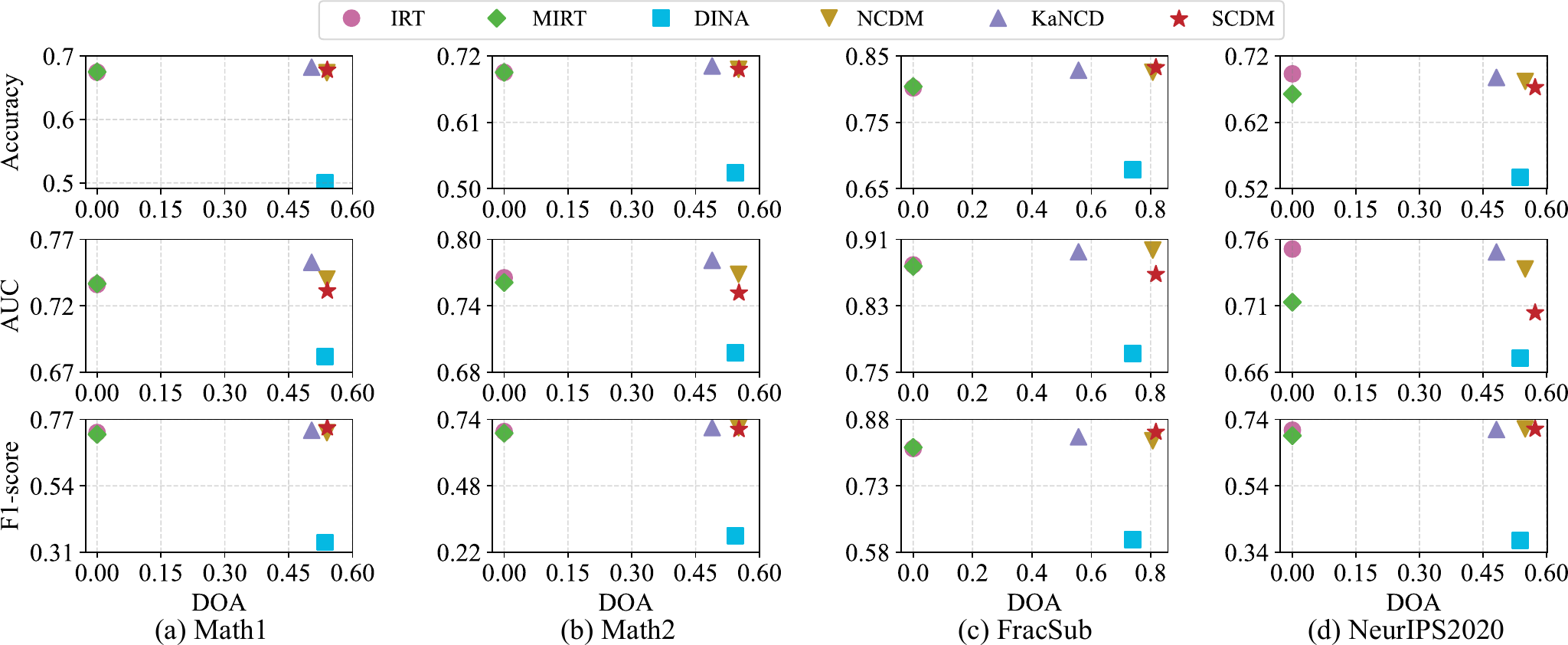}
  \caption{The Pareto performance of generalization and interpretability of SCDM and the compared methods}
  \label{fig::performance}
\end{figure*}

\textbf{Generalization Performance (Q1).} As shown in Figure~\ref{fig::performance}, SCDM performs competitively with the existing CDMs on each metric, without being dominated in any of them. It even outperforms other CDMs in terms of the F1-score. In fact, in the field of machine learning, enhancing model interpretability may weaken the generalization performance to a certain extent~\cite{du2020techniques}. DINA is an extreme case, with a high DOA but inadequate generalization performance. Therefore, demanding an improvement in generalization while increasing interpretability is quite challenging. Hence, SCDM can achieve improved interpretability of interaction function and diagnostic outcomes while maintaining strong generalization, indicating a well-balanced trade-off between these two aspects.

\textbf{Interpretability Performance (Q2).} We assess the interpretability of diagnostic outcomes. As shown in Figure~\ref{fig::performance}, the SCDM performs competitively or even outperforms all existing CDMs in terms of DOA, which shows SR also helps improve DOA. Notably, due to the utilization of latent vectors in IRT and MIRT, there does not exist an explicit correspondence between dimensions of latent vectors and knowledge attributes. Thus, we consider their DOA to be $0$  in Figure~\ref{fig::performance}. 

\begin{table*}[!htbp]
    \centering
    \caption{Ablation study of SCDM. ``SCDM w.o. GP'' refers to the SCDM without using GP, and ``SCDM w.o. Adam'' refers to the SCDM without using Adam to optimize parameters. In each column, an entry is marked in bold if its mean value is the best. By $t$-test, SCDM is significantly better than them on all the performance metrics with significant level $\alpha = 5\%$.}
    \resizebox{\linewidth}{!}{
    \begin{tabular}{l|cccc|cccc|cccc|cccc}
    \toprule
 & \multicolumn{4}{c|}{Math1} & \multicolumn{4}{c|}{Math2} & \multicolumn{4}{c|}{FracSub} & \multicolumn{4}{c}{NeurIPS2020} \\
    \midrule
 & Accuracy & AUC & F1 & DOA & Accuracy & AUC & F1 & DOA & Accuracy & AUC & F1 & DOA & Accuracy & AUC & F1 & DOA \\
    \midrule
SCDM w.o. GP & 67.22 & 72.46 & 73.95 & 53.13 & 69.05 & 75.07 & 69.94 & 54.92 & 82.67 & 86.33 & 84.33 & 81.16 & 63.43 & 65.98 & 67.50 & 56.03 \\
SCDM w.o. Adam & 67.30 & 72.86 & 73.81 & 50.97 & 68.43 & 75.08 & 68.67 & 51.58 & 76.42 & 83.12 & 78.62 & 68.68 & 63.27 & 68.70 & 68.62 & 51.37 \\
    \midrule
SCDM & \textbf{67.78} & \textbf{73.13} & \textbf{74.14} & \textbf{53.75} & \textbf{69.79} & \textbf{75.17} & \textbf{70.15} & \textbf{55.08} & \textbf{83.26} & \textbf{86.80} & \textbf{85.15} & \textbf{81.78} & \textbf{67.25} & \textbf{70.48} & \textbf{71.11} & \textbf{57.32} \\
\bottomrule
\end{tabular}
}
\label{tab::ablation}
\end{table*}

\textbf{Ablation Study (Q3).} In order to comprehend the impact of the GP and Adam components on SCDM's performance (expressed in percentage), an ablation study was conducted. The results shown in Table~\ref{tab::ablation}, where the ``SCDM w.o. GP'' indicates SCDM solely employing the initial interaction function. ``SCDM w.o. Adam'' signifies the utilization of a derivative-free evolutionary strategy~(see Appendix~\ref{para::woadam} for details) instead of the gradient-based Adam. By $t$-test, SCDM is significantly better than them on all the performance metrics with significant level $\alpha = 5\%$, and the variance of each metric is less than $0.05$, revealing that both components have a positive impact on the performance. Only using manually designed interaction functions may reduce SCDM's performance, but updating interaction functions with GP improves learning parameters, resulting in a performance enhancement. Besides, compared with the evolutionary strategy, the Adam optimizer makes learning parameters quicker and more accurate since it strictly adheres to the monotonicity assumption, leading to a significant enhancement in performance.

\begin{figure}[!t]
  \centering
  \includegraphics[width=0.60\linewidth]{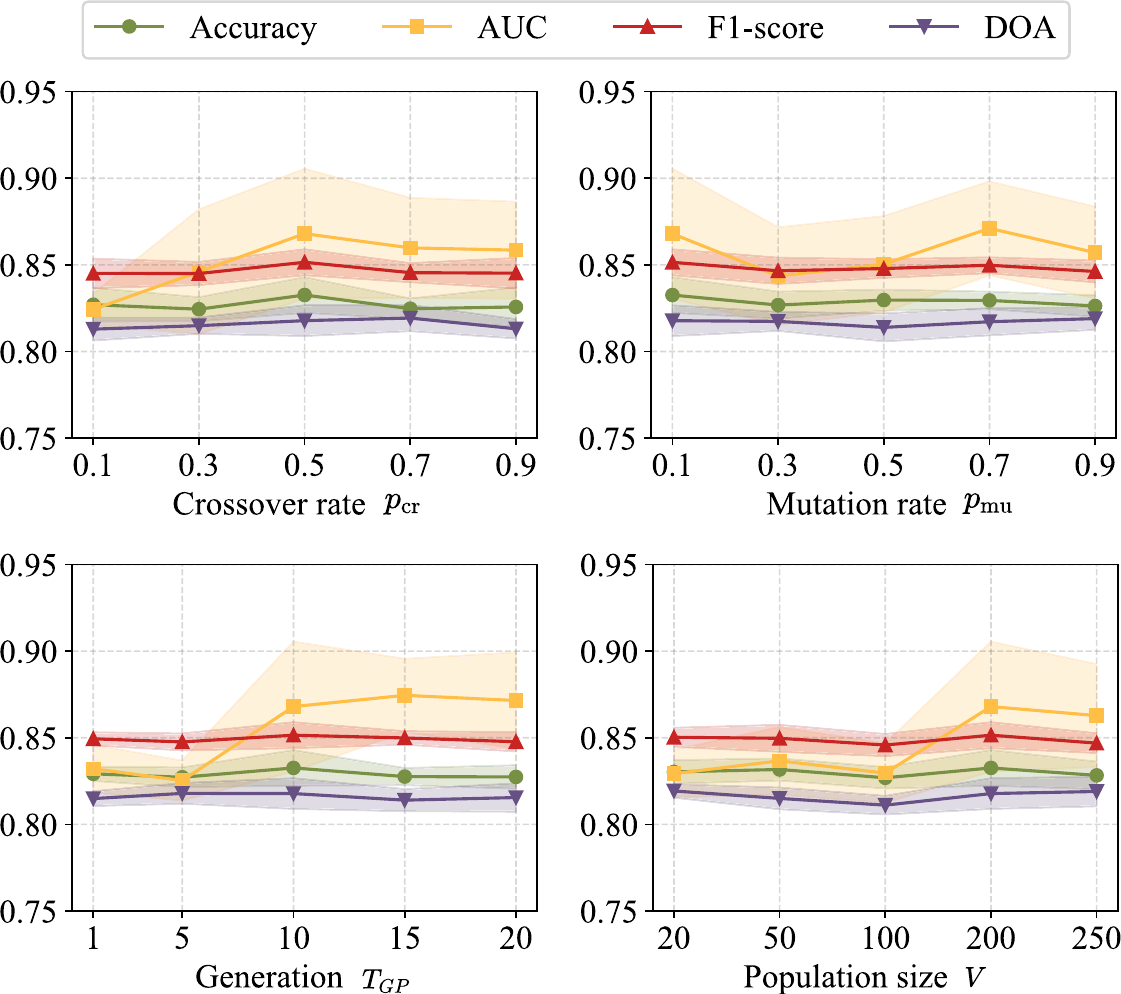}
  \caption{The performance of generalization and interpretability under different $p_{\text{cr}}$, $p_{\text{mu}}$, $T_{GP}$ and $V$ values on FracSub}
  \label{fig::hyper}
\end{figure}

\textbf{Hyperparameter Analysis (Q4).} We conduct a hyperparameter experiment to study the effect of crossover rate, mutation rate, generations and population size on the FracSub dataset whose results are similar to other datasets. Figure~\ref{fig::hyper} illustrates that our hyperparameter settings are good for most metrics. In practice, opting for a lower mutation rate and higher crossover rate promotes evolution, as the latter effectively blends various operators. A larger population size and more generations benefit evolution, but this entails a trade-off between performance and time. Besides, accuracy, F1-score and DOA are relatively insensitive to hyperparameters, whereas AUC shows higher sensitivity to them. Hence, when computational resources allow, opting for a larger population size and generations, coupled with suitable mutation and crossover rates, enhances the SCDM's performance.

\begin{figure}[!b]
  \centering
  \includegraphics[width=0.78\linewidth]{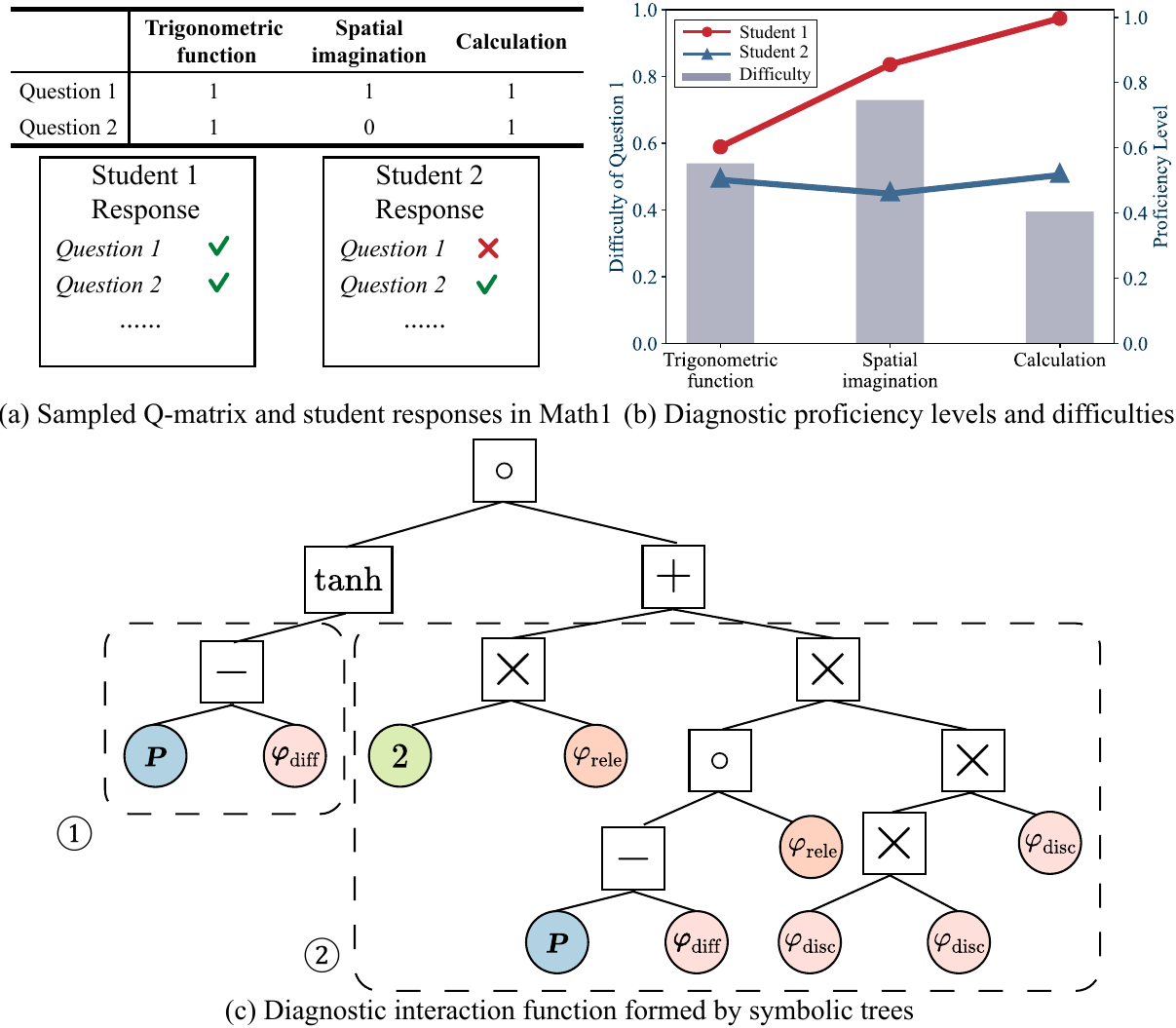}
  \caption{Case study: diagnostic outcomes of two students, exercise features and interaction function}
  \label{fig::case}
\end{figure}
\textbf{Case Study (Q5).} Since each obtained symbolic tree is unique and there is no true function available as a standard, it is challenging to quantify the interpretability of the symbolic tree. In this part, we delve into a more detailed analysis of interpretability. 
Suppose an educator intends to diagnose the cognitive state of students participating in the test of dataset Math1, one can employ SCDM for cognitive diagnosis of the students, and the outcomes are presented in the Figure~\ref{fig::case}. The bars in~(b) represent the difficulty on each knowledge attribute of Question 1, and the lines represent proficiency levels of Student 1 and Student 2 respectively. Figure~\ref{fig::case}~(b) depicts the necessity for students' proficiency levels to surpass the exercise difficulty levels to answer correctly, which is proved by the response logs in~(a). Besides, the interpretability of the interaction function shown in~(c) is noteworthy. Specifically, the tree depth is low and the choice of function set is highly interpretable since it satisfies the monotonicity assumption in education. For each part, Part~\ding{172} in the symbolic tree involves proficiency level calculations, where deducing exercise difficulty from proficiency levels is corresponding to the findings in (a) and (b), and it is reasonable in education. Part~\ding{173} contains complex computations concerning exercise discrimination and exercise-knowledge relevance. To be specific, 
the relevance $\bm{\varphi}_{\text{rele}}$ multiplied by discrimination suggests some underlying interconnections between knowledge attributes. These two parts result in the final interaction function.

\section{Conclusion}
This paper aims to simultaneously boost interpretability of outcomes and interaction functions, while maintaining competitive generalization performance. The proposed symbolic cognitive diagnosis incorporates the symbolic tree to explicably represent the complicated student-exercise interaction function and gradient-based optimization methods to effectively learn the student and exercise parameters. To effectively tunnel the discrete symbolic representation and continuous parameter optimization, and fulfill the SCD framework, we propose to hybridly optimize the representation and parameters in an alternating way. SCD possesses the merits of high intelligibility, generalization and flexibility. We sincerely hope this tentative work could pave the way for landing CDA in intelligent education systems. The future work of SCD includes theoretically disclosing the convergence behavior, further enhancing the generalization stability and exploring more intelligent education applications.

\section*{Acknowledgments}
The authors would like to express our sincere thanks to the anonymous reviewers for their constructive and valuable comments. The authors also would like to thank Hengzhe Zhang for the helpful discussion, and Xinyu Shi for the reliable support. The algorithms and datasets in this paper do not involve any ethical issue. This work is supported by the National Natural Science Foundation of China (62106076), Defense Industrial Technology Development Program (JCKY2019204A007), National Natural Science Foundation of China (92270119), and Natural Science Foundation of Shanghai (21ZR1420300).

\appendix
\section*{Appendix}
\renewcommand{\thealgorithm}{\Roman{algorithm}}
\renewcommand{\thetable}{\Roman{table}}
\renewcommand{\thefigure}{\Roman{figure}}

\setcounter{algorithm}{0}
\setcounter{table}{0}
\setcounter{figure}{0}

\section{Cognitive Diagnosis Assessment}
Cognitive Diagnosis Assessment (CDA) is a fundamental and crucial task in intelligent education systems. An illustrative example of CDA is shown in Figure~\ref{fig::CDAExample} of Appendix. There are three main factors in CDA: students, exercises and knowledge attributes which are also referred to as skills (e.g., calculation). The purpose of CDA is to model the student-exercise interaction via an interaction function based on response logs, and diagnose the students' cognitive states, i.e., inferring the proficiency levels on knowledge attributes.
\begin{figure*}[!h]
  \centering
  \includegraphics[width=0.92\linewidth]{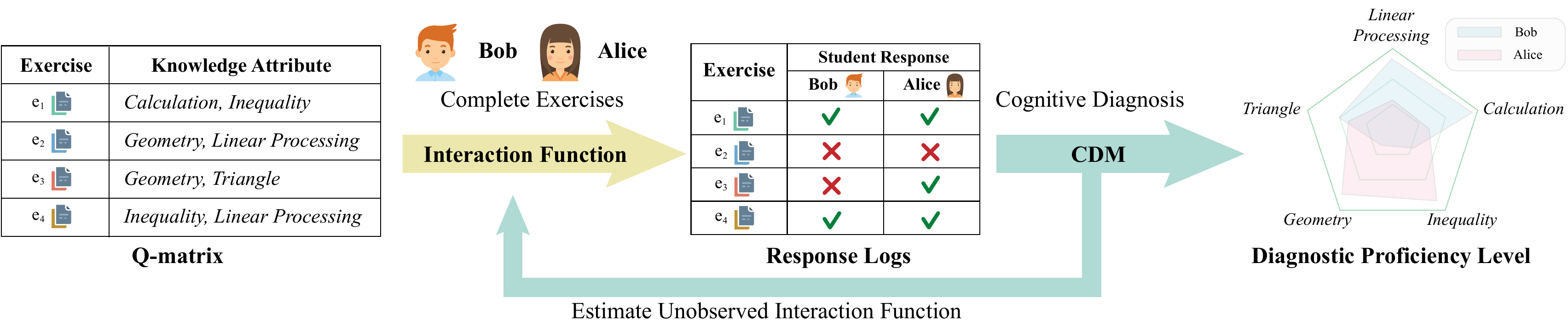}
  \caption{An illustrative example of cognitive diagnosis assessment}
  \label{fig::CDAExample}
\end{figure*}

\section{Details of Genetic Programming}\label{para::gpsetting}
In the SCD Model Implementation Section of the main paper, we have introduced the genetic programming to implement symbolic regression. Herein, we will list the details in Algorithm~\ref{alg::scdm} of the main paper.

\subsection{Detailed Settings}
We provide more detailed settings of implementation and experiments.

\textbf{Function Set.} In order to enhance the fitting capacity, the function set should encompass various symbols. For example, $\sin$ and $\cos$ can be involved in it in certain educational scenarios~\cite{fan2023integration}. However, to meet Assumption~\ref{def::ma}, these non-monotonic operators cannot be employed. Therefore, we define the function set as $\{+, \times, \circ, -, \tanh, {\rm{sigmoid}}\}$.

\textbf{Constraints of Symbolic Tree.} Recalling Eq.~\eqref{eq::if}, we notice that the output of an interaction function should be a scalar. However, this requirement cannot be satisfied by every symbolic tree generated. Therefore, certain constraints need to be imposed on symbolic trees. First, the input of a symbolic tree must include $\bm{P}_{i, \bullet}$ and $\varphi_{\text{rele}_j}$ to ensure that the final output is relevant to the knowledge attributes. Second, the computation for each operator should be valid (e.g., a unary operator only accepts one parameter). Finally, the output of a valid symbolic tree must be scalar. Therefore, after the creation of each symbolic tree (including the initialization and the genetic operations such as crossover, mutation and selection), it needs to be checked to ensure that it satisfies the aforementioned constraints.

\textbf{Fitness.} Accuracy is chosen as the metric of fitness. Similar to the way of measuring the generalization of SCDM, we assess the fitness of each individual on the training set of students' correctness. Special attention should be paid to that we do not use the test set when evaluating the fitness of each individual (i.e., symbolic tree). This is crucial during the training process to avoid the leakage problem.

\begin{algorithm}[!b]
\caption{Crossover}
\label{alg::crossover}
\textbf{Input}: Two individuals $f_1$, $f_2$.\\
\textbf{Output}: New individuals $f_1'$, $f_2'$.
\begin{algorithmic}[1] 
\IF {${\rm{height}}(f_1) < 2$ or ${\rm{height}}(f_2) < 2$}
\STATE \textbf{return} $f_1$, $f_2$
\ELSE
\STATE $f_1', f_2' \gets f_1, f_2$
\STATE randomly choose nodes $node_1$, $node_2$ from $f_1'$, $f_2'$ respectively
\STATE search the subtrees $\tilde{f}_1$, $\tilde{f}_2$ beginning from $node_1$, $node_2$ respectively
\STATE $\tilde{f}_1, \tilde{f}_2 \gets \tilde{f}_2, \tilde{f}_1$
\STATE \textbf{return} $f_1'$, $f_2'$
\ENDIF
\end{algorithmic}
\end{algorithm}

\begin{algorithm}[!t]
\caption{Mutation}
\label{alg::mutation}
\textbf{Input}: Individuals $f$.\\
\textbf{Output}: New individual $f'$.
\begin{algorithmic}[1] 
\STATE $f' \gets f$
\IF{${\rm{uniform}(0, 1)} < 0.5$}
\STATE randomly choose node $node_1$ from $f'$
\STATE search the subtree $\tilde{f}_1$ beginning from $node_1$
\STATE randomly generated a subtree $\tilde{f}$
\STATE let $node_1$ and its subtree $\tilde{f}_1$ become left child of $\tilde{f}$
\STATE $\tilde{f}_1 \gets \tilde{f}$
\ELSE
\IF {${\rm{height}}(f) < 5$}
\STATE \textbf{return} $f$
\ELSE
\STATE randomly choose node $node_1$ from $f'$
\STATE search the subtree $\tilde{f}_1$ beginning from $node_1$
\STATE randomly choose node $node_2$ from $\tilde{f}_1$
\STATE search the subtree $\tilde{f}_2$ beginning from $node_2$
\STATE $\tilde{f}_1 \gets \tilde{f}_2$
\ENDIF 
\ENDIF
\STATE \textbf{return} $f'$
\end{algorithmic}
\end{algorithm}

\subsection{Algorithmic Details}
We elaborate the details in Algorithm~\ref{alg::scdm} that are sketched in the main paper due to the page limitation.

\textbf{$\mathbf{uniform(0, 1)}$.} To simulate the random genetic operations (crossover, mutation and selection), we introduce a uniform distribution. In this paper, unless otherwise specified, the random numbers used are uniformly distributed between $0$ and $1$, denoted as ${\rm{uniform}(0, 1)}$.

\textbf{Crossover.} We use an intuitive and simple way to realize crossover: randomly select crossover point in each individual and exchange each subtree with the point as root between each individual~\cite{o2009riccardo}. It is shown in Algorithm~\ref{alg::crossover} of Appendix.

\textbf{Mutation.} The mutation involves two parts: insert and prune. The former enhances tree complexity, while the latter prevents excessive complexity. The detail implementation is shown in Algorithm~\ref{alg::mutation} of Appendix. From line~3 to 7 in Algorithm~\ref{alg::mutation}, a new branch will be inserted at a random position in individual, and this subtree at the chosen position is used as child node of the created subtree. From line~9 to 16, a branch will be randomly chosen and replaced with one of the branch's nodes~\cite{deap}.

\textbf{Selection.} The selection is implemented by tournament selection~\cite{deap}, which selects the best individual among population $T_{\text{se}}$ times and gives the selected individuals. Algorithms~\ref{alg::selection} in Appendix shows the details. In the implementation of GP, $k$ is set as $3$.

\begin{algorithm}[!t]
\caption{Tournament Selection}
\label{alg::selection}
\textbf{Input}: Population $\{f_1, f_2, \dots, f_V\}$, selection times $T_{\text{se}}$.\\
\textbf{Output}: Selected individuals $\{f_1, f_2, \dots, f_{T_{\text{se}}}\}$.

\begin{algorithmic}[1] 
\STATE initialize selected individual list $l \gets \{\}$
\WHILE{${\rm{length}}(l) < T_{\text{se}}$}
\STATE obtain $f'$ from $\{f_1, f_2, \dots, f_V\}$, where $f'$ is the fittest individual in the population
\STATE add $f'$ to $l$
\STATE remove $f'$ from population
\ENDWHILE
\STATE \textbf{return} $l$
\end{algorithmic}

\end{algorithm}

\section{Extra Experiment Results}
\label{para::exp}

\subsection{Statistics of SCDM Performance}
\label{para::stat}
All statistics in the generalization performance and interpretability performance are shown in Table~\ref{tab::performance} (expressed in percentage) of Appendix. Since IRT and MIRT utilize latent vectors, DOA is unavailable, and thus we use ``---'' to represent it. SCDM outperforms existing models on half of metrics of each dataset and performs competitively with other CDMs on the left metrics. The interpretability metric (DOA) of SCDM is always the best in Table~\ref{tab::performance}.

\begin{table*}[!b]
    \centering
    \caption{Generalization and interpretability performance (expressed in percentage) of baselines, state-of-the-art methods and SCDM. In each column, an entry is marked in bold if its mean value is the best and underline for the runner-up. By $t$-test, a bold one is significantly better than others on the corresponding metrics with significant level $\alpha = 5\%$. The symbol ``---'' means that the value is unavailable.}
    \resizebox{\linewidth}{!}{
    \begin{tabular}{c|cccc|cccc|cccc|cccc}
    \toprule
 & \multicolumn{4}{c|}{Math1} & \multicolumn{4}{c|}{Math2} & \multicolumn{4}{c|}{FracSub} & \multicolumn{4}{c}{NeurIPS2020} \\
    \midrule
 & Accuracy & AUC & F1 & DOA & Accuracy & AUC & F1 & DOA & Accuracy & AUC & F1 & DOA & Accuracy & AUC & F1 & DOA \\
    \midrule
IRT & 67.44 & 73.61 & 72.49 & --- & 69.28 & 76.54 & 69.27 & --- & 80.19 & 87.94 & 81.44 & --- & \textbf{69.31} & \textbf{75.28} & 70.79 & --- \\ 
MIRT & 67.49 & 73.67 & 71.83 & --- & 69.28 & 76.11 & 68.64 & --- & 80.39 & 87.77 & 81.69 & --- & 66.27 & 71.28 & 69.12 & --- \\ 
DINA & 50.00 & 68.20 & 34.43 & 53.61 & 52.68 & 69.78 & 28.55 & 54.31 & 67.82 & 77.24 & 60.81 & 73.89 & 53.75 & 67.08 & 37.52 & 53.85 \\ 
NCDM & 67.36 & \underline{74.02} & 72.43 & \underline{53.67} & 69.78 & \underline{76.85} & \underline{70.78} & \underline{55.01} & 82.55 & \textbf{89.72} & 83.19 & \underline{80.72} & 68.17 & 73.78 & \underline{71.06} & \underline{54.98} \\ 
KaNCD & \textbf{68.23} & \textbf{75.28} & \underline{73.27} & 50.40 & \textbf{70.35} & \textbf{78.12} & \textbf{70.84} & 48.88 & \underline{82.85} & \underline{89.51} & \underline{84.08} & 55.69 & \underline{68.77} & \underline{75.06} & 70.98 & 48.19 \\
SCDM & \underline{67.78} & 73.13 & \textbf{74.14} & \textbf{53.75} & \underline{69.79} & 75.17 & 70.15 & \textbf{55.08} & \textbf{83.26} & 86.80 & \textbf{85.15} & \textbf{81.78} & 67.25 & 70.48 & \textbf{71.11} & \textbf{57.32} \\ 
\bottomrule
\end{tabular}
}
\label{tab::performance}
\end{table*}

\begin{algorithm}[!t]
\caption{Evolutionary Strategy (ES)}
\label{alg::es}
\textbf{Input}: Response logs $R=\{s, e, r\}$, Q-matrix: $\bm{Q}$, current interaction function $f_{\text{curr}}$.\\
\textbf{Parameter}: generation $T_{\text{ES}}$, population size of ES $V_{\text{ES}}$, mutation rate of ES $p_{\text{es-mu}}$, learning rate $lr$.\\
\textbf{Output}: Proficiency levels $\bm{P}$, exercise features $\bm{\varphi}_{\text{diff}}, \varphi_{\text{disc}}$.

\begin{algorithmic}[1] 
\STATE Randomly initialize a population of parameters ${\rm{Para}}=$\\
\resizebox{\linewidth}{!}{
$\{\{\bm{P}^{(1)}, \bm{\varphi}_{\text{diff}}^{(1)}, \varphi_{\text{disc}}^{(1)}\}, \{\bm{P}^{(2)}, \bm{\varphi}_{\text{diff}}^{(2)}, \varphi_{\text{disc}}^{(2)}\}, \dots, \{\bm{P}^{(V_{\text{ES}})}, \bm{\varphi}_{\text{diff}}^{(V_{\text{ES}})}, \varphi_{\text{disc}}^{(V_{\text{ES}})}\}\}$
}
\FOR {$t_1 = 1, 2, \dots, T_{\text{ES}}$}
\FOR {$i = 1, 2, \dots, V_{\text{ES}}$}
\IF {${\rm{uniform}}(0, 1) < p_{\text{es-mu}}$}
\STATE $\bm{P}^{(i)} \gets \bm{P}^{(i)} + lr \cdot {\rm{normal}}(0, 1)$
\STATE $\bm{\varphi}_{\text{diff}}^{(i)} \gets \bm{\varphi}_{\text{diff}}^{(i)} + lr \cdot {\rm{normal}}(0, 1)$
\STATE $\varphi_{\text{disc}}^{(i)} \gets \varphi_{\text{disc}}^{(i)} + lr \cdot {\rm{normal}}(0, 1)$
\ENDIF
\ENDFOR
\STATE ${\rm{Para}} \gets \text{selection}({\rm{Para}})$
\STATE Evaluate all individuals in ${\rm{Para}}$, 
\ENDFOR
\STATE $\bm{P}, \bm{\varphi}_{\text{diff}}, \varphi_{\text{disc}} \gets \bm{P}', \bm{\varphi}_{\text{diff}}', \varphi_{\text{disc}}'$, where $\bm{P}', \bm{\varphi}_{\text{diff}}', \varphi_{\text{disc}}'$ are the fittest parameters in $\rm{Para}$
\STATE \textbf{return} $\bm{P}, \bm{\varphi}_{\text{diff}}, \varphi_{\text{disc}}$
\end{algorithmic}
\end{algorithm}

\subsection{SCDM without (w.o.) Adam}
\label{para::woadam}
In order to implement parameter optimization without utilizing the Adam optimizer, we introduce the evolutionary strategy (ES)~\cite{beyer2002evolution} for optimizing the object function as Eq.~\eqref{appe::eq::loss} to obtain the optimal parameters. 
\begin{equation}
\label{appe::eq::loss}
\mathcal{L}_{r\in R}(y_f, r) = -\sum\limits_{i=1}^{|R|}(r^{(i)}\log y_f^{(i)} + (1-r^{(i)})\log (1-y^{(i)}))\,,
\end{equation}
where $y_f^{(i)}$ is the $i$-th output of interaction function $f$ and $r^{(i)}$ is the corresponding label. Algorithm~\ref{alg::es} in Appendix describes the detailed implementation.

Algorithm~\ref{alg::es} is similar to the GP module of aforementioned Algorithm~\ref{alg::scdm} in the main paper. However, to better suit the ES, certain modifications have been introduced. Specifically, the learning rate is $0.1$, the population size is $100$, the number of generation is $50$, and the mutation rate is $0.5$ which is much larger than that of GP due to full coverage the search space. The fitness of each individual is the value of loss (the smaller the better). ${\rm{normal}}(0, 1)$ is a normally distributed random variable with zero mean and standard deviation $1$. The selection is as same as Algorithm~\ref{alg::selection} in Appendix, and $k = 3$.

\clearpage
\bibliographystyle{named}
\bibliography{main}

\end{document}